\documentclass[12pt]{article}
\usepackage{graphicx}
\usepackage{amsmath}
\usepackage{geometry}
\usepackage{booktabs}
\usepackage{hyperref}
\usepackage{caption}
\usepackage{float}
\usepackage{cite}
\title{Detection of Temporal Variability in U.S. Climate Using Harmonic and Wavelet Decomposition}
\author{Thomas Xiao}
\date{\today}

\begin{document}
\maketitle

\begin{abstract}
This study investigates temporal variability in U.S. climate using harmonic decomposition techniques, specifically Fourier and wavelet transforms. 
Monthly temperature, precipitation, and drought index data from the National Oceanic and Atmospheric Administration (NOAA) U.S. Climate Divisional Dataset (nClimDiv, 1895–2024) were analyzed to detect periodic structures and their evolution over time. 
By comparing harmonic-based models with linear regression trends, this research evaluates the explanatory power of cyclic components in reproducing and predicting observed variability. 
Results show that U.S. climate records exhibit dominant periodicities near one year (seasonal) and 2–7 years (associated with the El Niño–Southern Oscillation, ENSO), and that incorporating harmonic terms significantly improves model performance across most states and variables. 
The findings indicate that U.S. climate fluctuations are characterized by quasi-stationary oscillations rather than purely monotonic trends. \textbf{Overall, the main implication is that frequency-aware models provide measurably better predictive skill than trend-only approaches and should be incorporated into seasonal outlooks, drought monitoring, and resource planning.}
\end{abstract}

\section{Introduction}
Quantifying long-term climate change has often relied on estimating linear trends in temperature, precipitation, and related indices. Such approaches are widespread and useful for summarizing monotonic change at regional to global scales (e.g., trend reviews and applications in climatology) \cite{Mudelsee2019Review,IPCCAR6Ch1,Santer2008Trends}. However, short-to-intermediate time-scale variability—especially seasonal and interannual oscillations—also governs practical risks in agriculture, water resources, and hazard management. Temperature and precipitation do not change uniformly through time; their timing and magnitude fluctuate through recurring cycles, including the annual cycle and interannual variability associated with El Niño–Southern Oscillation (ENSO), as well as lower-frequency decadal variability. In operational contexts, the timing of precipitation or heat episodes can be as consequential as their mean change; for example, shifts in seasonal onset affect planting, irrigation scheduling, and reservoir operations. Historical episodes such as the 1930s Dust Bowl and high-amplitude El Niño events (1982–1983, 1997–1998) demonstrate that departures from the mean often arise from structured variability rather than smooth trends. While many studies emphasize linear trend estimation and its uncertainty (including the role of autocorrelation in inflating trend significance) \cite{Mudelsee2019Review}, there has been comparatively limited work that systematically evaluates the \emph{predictive contribution of harmonic components}—identified from frequency-domain analysis—across U.S. climate variables and states, and contrasts that performance against trend-only baselines. This study addresses that gap by (i) extracting dominant periodicities with Fourier analysis, (ii) assessing their time dependence with wavelets, and (iii) testing whether a regression that augments a linear trend with selected harmonic terms improves fit and near-term forecasts relative to a trend-only model.

\section{Data and Preprocessing}
\subsection{Dataset}
The NOAA U.S. Climate Divisional Dataset (nClimDiv) provides monthly climate records for all states and climate divisions from 1895 to the present. 
Data are derived from the Global Historical Climatology Network–Daily (GHCN-D). Variables used in this analysis include:
\begin{itemize}
    \item Average, maximum, and minimum temperature (°F);
    \item Precipitation (inches);
    \item Drought indices: Palmer Drought Severity Index (PDSI), Palmer Modified Drought Index (PMDI), Palmer Hydrological Drought Index (PHDI), Z-Index (ZNDX), and Standardized Precipitation Index (SPI) for multiple accumulation periods.
\end{itemize}

\subsection{Anomaly Computation}
Because mean climate conditions vary substantially by location, all series were converted to standardized anomalies relative to the 1981–2010 climatological baseline. For instance, a two-inch rainfall may represent an extreme event in arid regions such as Arizona but a typical monthly value in Florida. To account for this difference, the monthly anomaly $A_t$ is defined as:
\[
A_t = X_t - \bar{X}_m,
\]
where $X_t$ is the observed monthly value and $\bar{X}_m$ is the long-term mean for that calendar month. This process removes the recurring seasonal signal, isolating deviations attributable to interannual or decadal variability.

\begin{figure}[H]
\centering
\includegraphics[width=\textwidth]{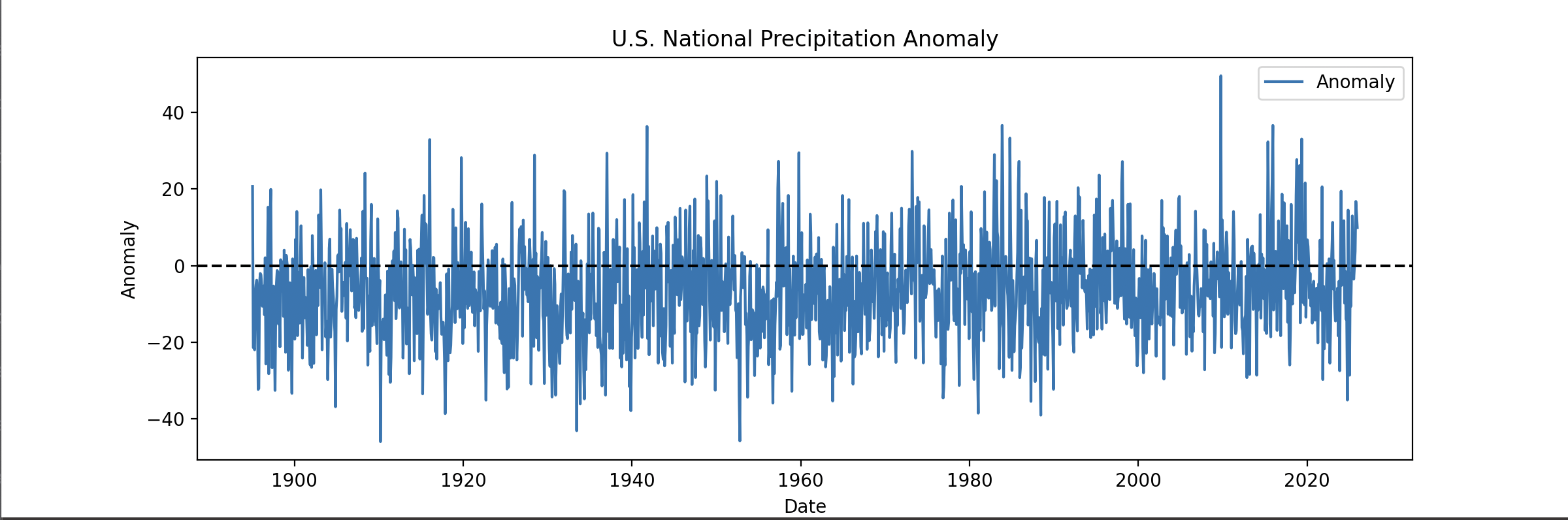}
\caption{Time series of U.S. national precipitation anomalies, 1895–2024.}
\end{figure}
The anomaly series (Figure 1) highlights both short-term and long-term departures from mean conditions. 
Negative anomalies during the 1930s correspond to the Dust Bowl drought, while strong positive anomalies around 1982–1983 and 1997–1998 align with major El Niño episodes. These associations confirm that anomalies effectively capture climate variability related to large-scale ocean–atmosphere processes. Other large positive and negative anomalies outside the Dust Bowl and major El Niño periods can be explained by a combination of widespread hydrologic extremes and data-related factors.
Negative departures during the 1950s correspond to the extended drought affecting the Southern Plains and Southwest, while weaker dry phases in the 1970s coincide with the 1976–1977 Pacific regime shift.
Large positive anomalies in the 1970s, early 1980s, and mid-2010s reflect nationwide wet periods linked to strong El Niño activity and enhanced subtropical moisture transport.
Some early-century spikes likely result from limited spatial sampling before network expansion, which caused regional events to disproportionately influence national means.

\section{Methods}
 We first identify dominant periodicities using the Discrete Fourier Transform (DFT), which quantifies which frequencies explain most variance. We then assess how those frequencies change over time using the Continuous Wavelet Transform (CWT), providing time–frequency localization. Next, we evaluate the predictive value of frequency information by comparing a trend-only regression to a harmonic hybrid regression that includes selected DFT components. Finally, we detect non-periodic structural changes using change-point analysis (PELT) to capture abrupt regime shifts that harmonics and trends do not represent.

\subsection{Fourier Transform}
The Discrete Fourier Transform (DFT) decomposes a signal into sinusoidal components at different frequencies:
\[
A_t = \sum_{k=1}^{N/2} [a_k \cos(2\pi f_k t) + b_k \sin(2\pi f_k t)],
\]
where $f_k$ is frequency in cycles per year, and $a_k$, $b_k$ represent the amplitude and phase of each frequency component. 
The resulting Fourier spectrum reveals the dominant periodicities contributing to observed variability.

\subsection{Wavelet Transform}
To address nonstationarity, the Continuous Wavelet Transform (CWT) captures time-varying frequency content:
\[
W(s, \tau) = \frac{1}{\sqrt{s}} \int A(t)\psi^*\!\left(\frac{t - \tau}{s}\right)dt,
\]
where $s$ denotes scale, $\tau$ represents time, and $\psi$ is the mother wavelet. 
This provides localized spectral information and identifies temporal shifts in periodic strength.

\begin{figure}[H]
\centering
\includegraphics[width=\textwidth]{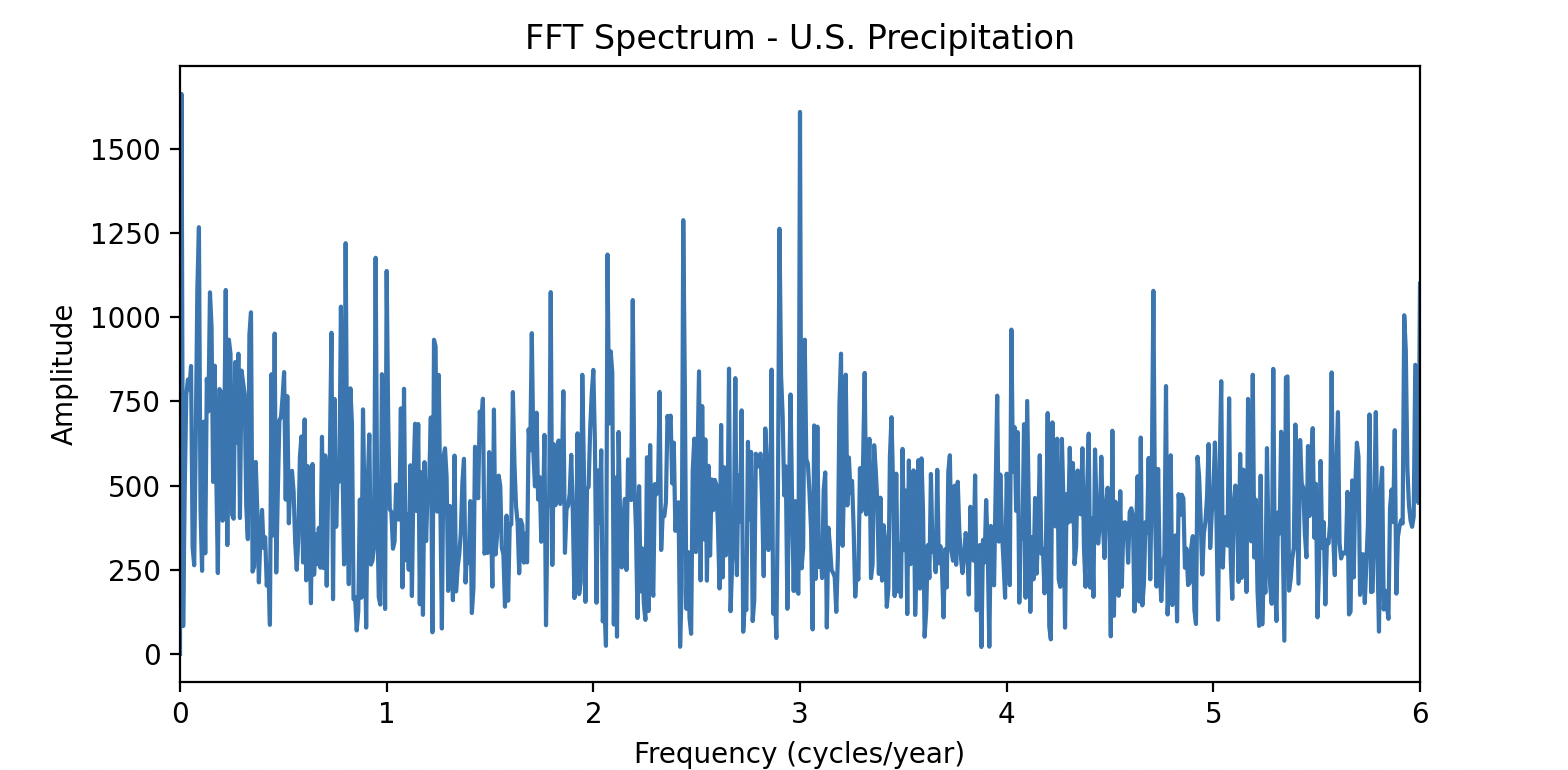}
\caption{Fourier amplitude spectrum of precipitation anomalies.}
\end{figure}

Figure 2 shows a dominant spectral peak near 3 cycles per year (approximately a 4-month period), with secondary energy at 2–4 cycles per year and comparatively weaker power near 1 cycle per year.
Because the analysis was performed on monthly anomalies with the seasonal mean removed, sub-annual harmonics dominate the variance.
The smaller feature near 1 cycle per year represents residual annual variability that persists due to incomplete removal of the seasonal cycle or non-stationary seasonality in precipitation patterns.
Overall, the spectral structure indicates that short-term (3–6-month) fluctuations contribute more strongly to precipitation variability than strictly annual or interannual components.

\begin{figure}[H]
\centering
\includegraphics[width=\textwidth]{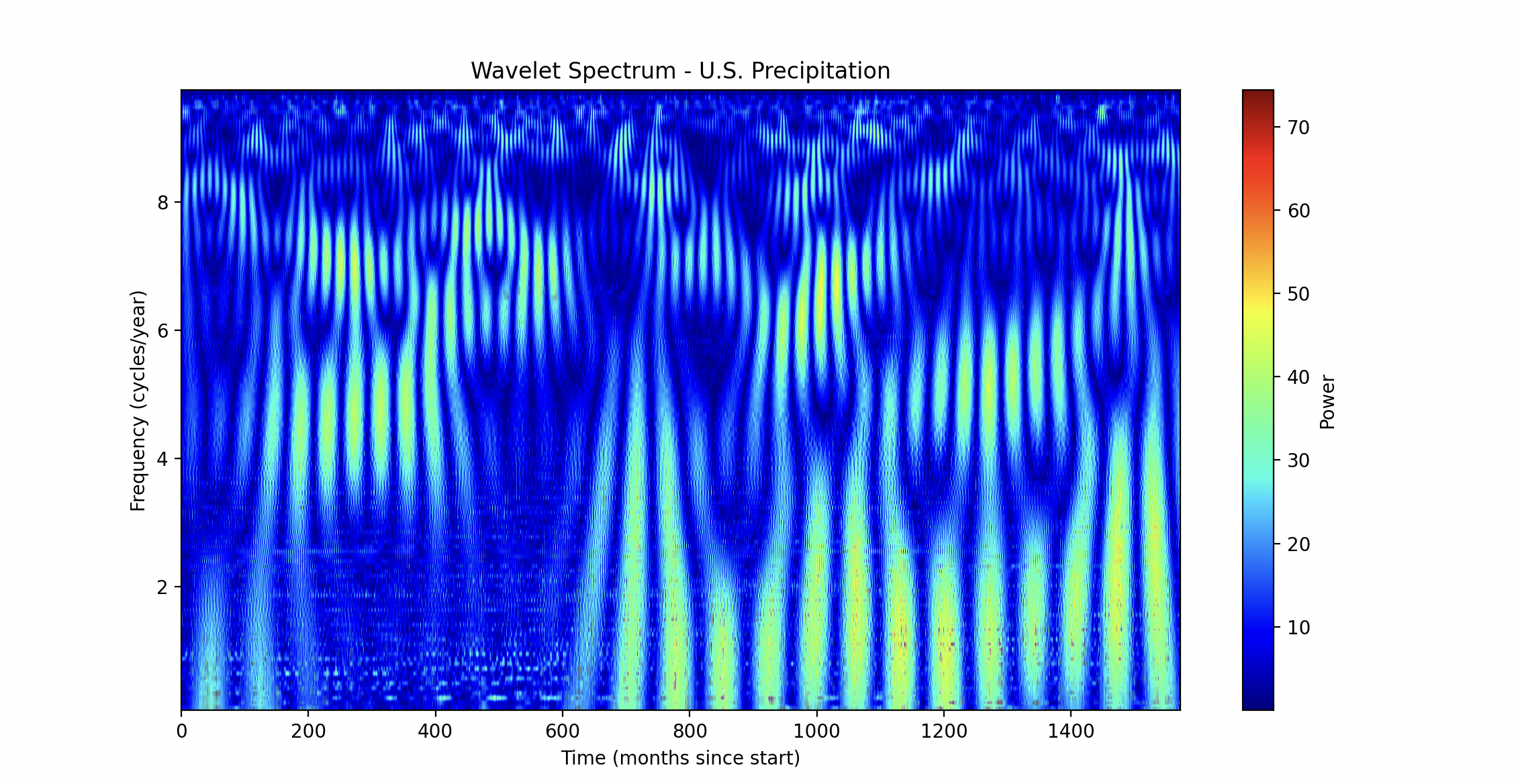}
\caption{Continuous wavelet power spectrum of precipitation anomalies.}
\end{figure}

In Figure 3, the wavelet power spectrum displays the evolution of frequency strength through time. 
The horizontal axis represents time (1895–2024), the vertical axis frequency in cycles per year, and the color scale indicates normalized wavelet power. Warm colors (yellow to red) denote strong periodic signals, while cool colors (blue) correspond to weak or absent oscillations. 
A continuous bright band near one cycle per year confirms that the annual precipitation cycle is persistent throughout the entire record. 
Enhanced power in the 2–7~year band between approximately 1950 and 2000 reflects stronger interannual variability associated with El~Niño–Southern~Oscillation (ENSO) activity, whereas weaker power at those scales in other periods indicates reduced ENSO influence. Faint power at longer periods (below 0.2~cycles~per~year) suggests the presence of decadal-scale oscillations, possibly related to broader Pacific climate modes. 
Overall, the spectrum illustrates that while the annual cycle is stable, interannual variability fluctuates in amplitude over time, demonstrating the value of wavelet analysis for time–frequency characterization.

\subsection{Trend and Harmonic Modeling}

To quantify both monotonic and cyclical components in the climate anomaly series, two regression frameworks were applied: a purely linear model and a harmonic hybrid model. 
The objective was to determine whether including periodic terms extracted from spectral analysis improves the representation and predictability of observed variability.
\\
\\
The \textbf{linear model} is formulated as:
\[
\hat{A}_t = \beta_0 + \beta_1 t + \epsilon_t,
\]
where $\hat{A}_t$ is the predicted anomaly at time $t$, $\beta_0$ is the intercept, $\beta_1$ is the linear trend coefficient, and $\epsilon_t$ represents independent, normally distributed residuals. 
This model assumes a constant rate of change over time and is equivalent to fitting a straight line through the anomaly record. 
While suitable for estimating long-term trends such as gradual warming or wetting, it is limited in its ability to represent recurring fluctuations or quasi-periodic oscillations that are common in climate systems.
\\
\\
To capture these oscillatory behaviors, a \textbf{harmonic hybrid model} was also employed:
\[
\hat{A}_t = \beta_0 + \beta_1 t + \sum_{k=1}^{K} [c_k \cos(2\pi f_k t) + d_k \sin(2\pi f_k t)] + \epsilon_t,
\]
where $f_k$ are the dominant frequencies identified through the Discrete Fourier Transform (DFT), and $c_k$ and $d_k$ are the corresponding amplitude coefficients. 
Each term in the summation represents a repeating wave component of frequency $f_k$, allowing the model to reproduce both the direction of long-term change and the recurrent variability observed in the data. 
\\
\\
In this framework, the linear term accounts for secular trends (e.g., steady increases in temperature or precipitation), while the harmonic terms describe cyclical variations such as the annual cycle, the 2–7~year ENSO band, and decadal oscillations. 
Model performance was evaluated using the \textbf{root mean square error (RMSE)}:
\[
\mathrm{RMSE} = \sqrt{\frac{1}{N} \sum_{t=1}^{N} (A_t - \hat{A}_t)^2},
\]
which quantifies average predictive deviation between modeled and observed anomalies. 
Residual diagnostics were also examined to verify that remaining errors were approximately white noise, confirming that major periodic and trend components had been adequately captured. This approach is particularly relevant to climate analysis because it distinguishes between slow, monotonic changes (for instance, the gradual increase in mean U.S. precipitation since the early 20th century) and recurrent fluctuations driven by natural variability. 
Comparing RMSE values between models demonstrates whether including cyclic structure yields a statistically and physically superior representation of observed patterns.

\subsection{Change-Point Detection}

To complement the regression analysis and identify non-periodic structural changes, the \textbf{PELT (Pruned Exact Linear Time)} algorithm was used to detect shifts in mean and variance across the anomaly series. 
Change-point detection identifies moments in time when the statistical properties of a signal—its average level or variability—undergo abrupt transitions. 
The PELT algorithm minimizes a penalized cost function:
\[
C(m) = \sum_{i=1}^{m+1} [\mathcal{L}(y_{(\tau_{i-1}+1):\tau_i})] + \beta m,
\]
where $\mathcal{L}$ is a segment-specific loss (here based on a radial basis function cost model), $\tau_i$ are the change-point indices, $m$ is the number of detected segments, and $\beta$ is a penalty parameter controlling model complexity. 
The algorithm efficiently determines both the optimal number and position of change points by pruning suboptimal solutions, achieving linear computational complexity with respect to series length. In the context of climate time series, change-point detection isolates major regime shifts such as the onset and termination of prolonged droughts or transitions between different ocean–atmosphere circulation states. 
For example, pronounced changes detected near the 1930s and 1970s correspond to the Dust Bowl drought and the Pacific climate regime shift, respectively. 
Unlike harmonic regression, which models smooth periodic variability, the PELT method captures discrete, non-repeating transitions indicative of abrupt climatic reorganization. 
Together, harmonic modeling and change-point detection provide a complementary framework: the former characterizes continuous oscillations, while the latter identifies structural discontinuities that define major epochs in U.S. climate history.

\section{Results and Discussion}
\begin{figure}[h!]
\centering
\includegraphics[width=\textwidth]{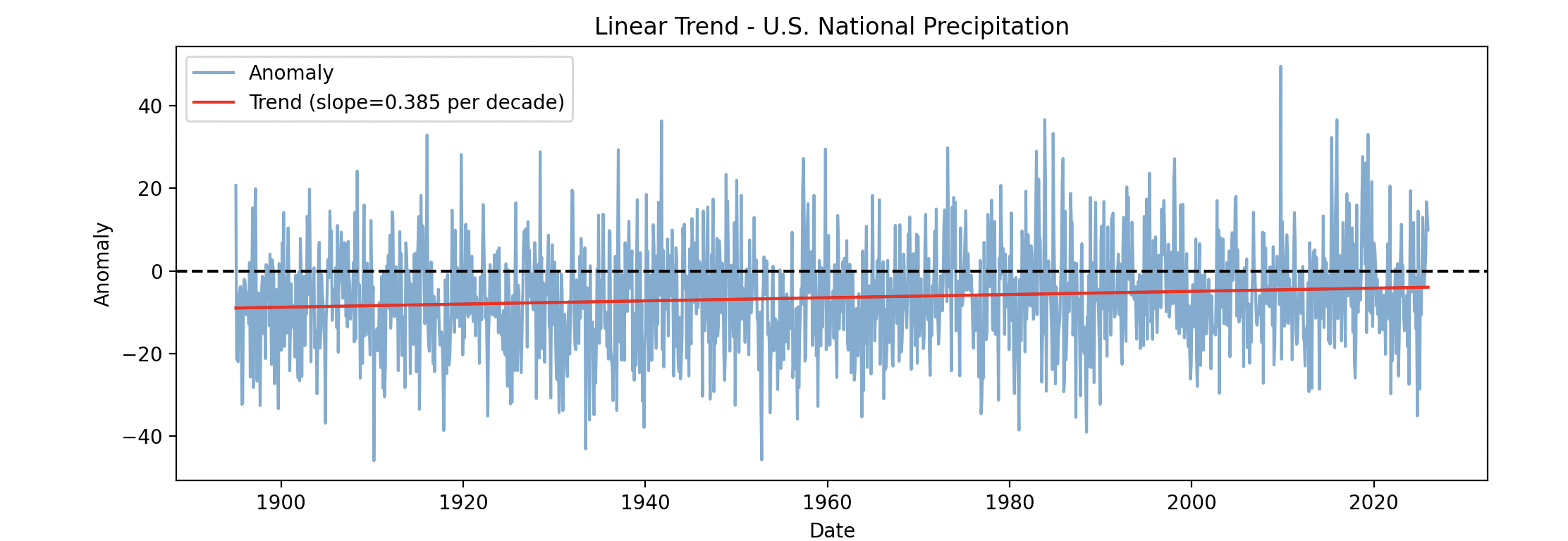}
\caption{Linear regression trend in U.S. precipitation anomalies.}
\end{figure}

Figure 4 shows a modest positive linear trend over the 1895–2024 period, consistent with a gradual increase in mean U.S. precipitation. 
However, interannual and decadal variability far exceed the linear component, confirming that trend-only models underrepresent short-term fluctuations.

\begin{figure}[H]
\centering
\includegraphics[width=\textwidth]{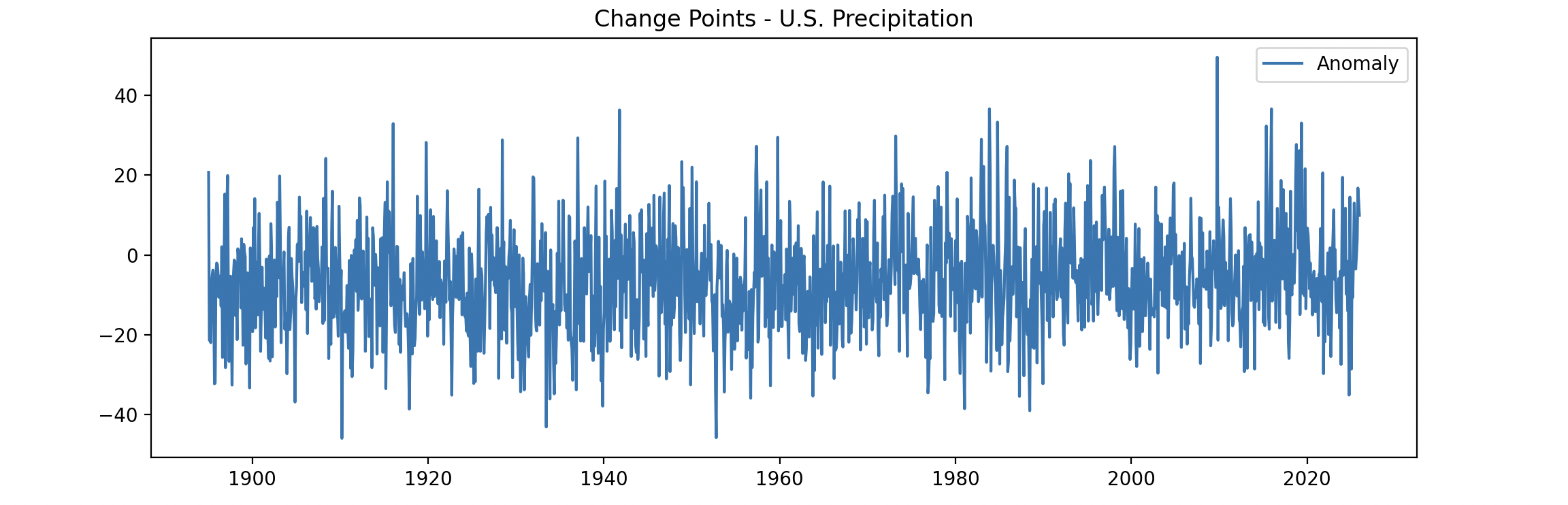}
\caption{Detected change points in U.S. precipitation anomalies.}
\end{figure}

In Figure 5, detected breakpoints align with major climatic regime transitions, including the 1930s drought and the 1976–1977 Pacific shift. 
These structural changes indicate that U.S. hydroclimate variability exhibits episodic reorganizations rather than gradual shifts.

\begin{figure}[H]
\centering
\includegraphics[width=0.9\textwidth]{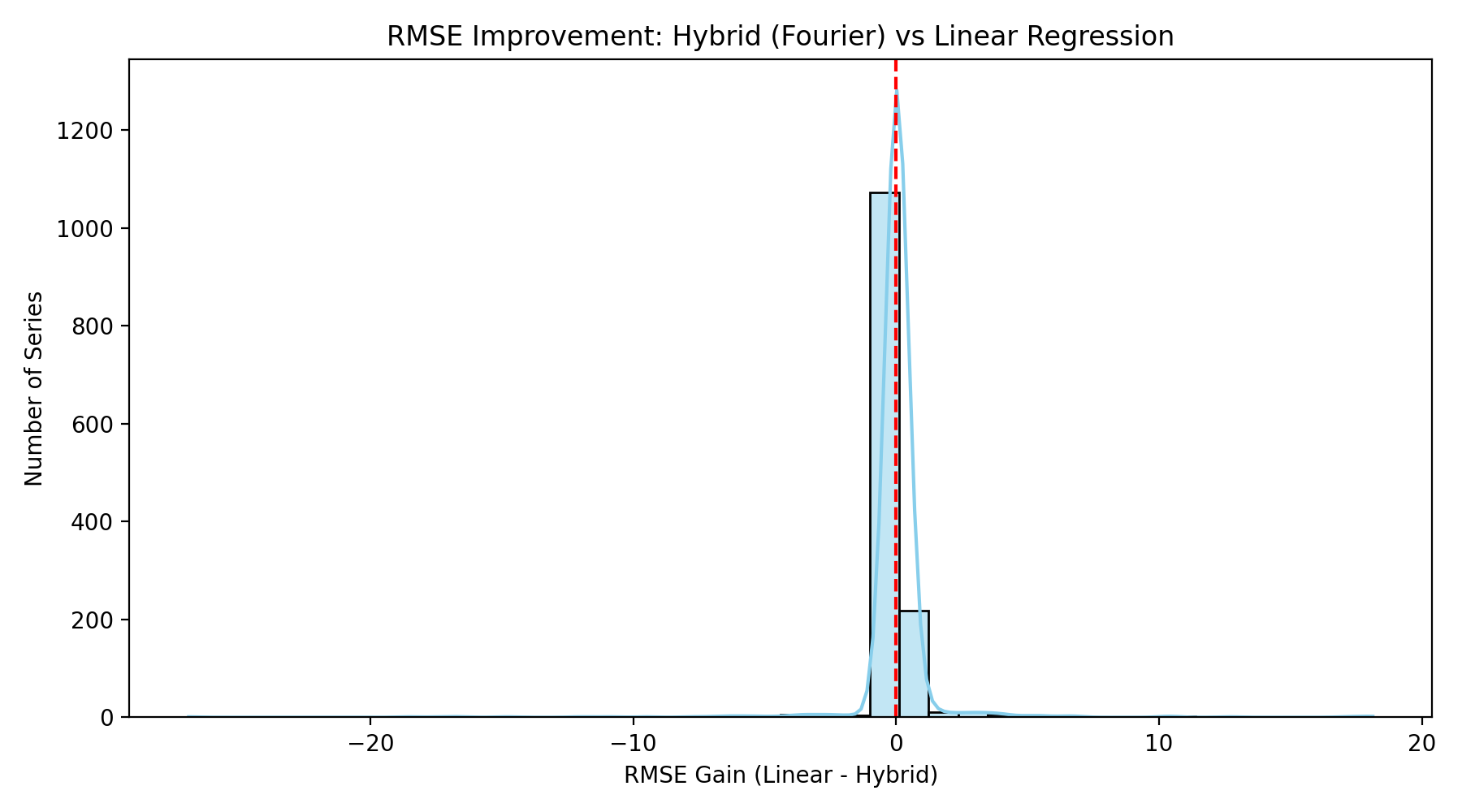}
\caption{Distribution of RMSE improvement for harmonic versus linear models.}
\end{figure}

As shown in Figure 6, harmonic models produce lower RMSE values for most states and variables. 
This demonstrates that periodic components explain a substantial proportion of variance beyond that captured by linear trends.

\begin{figure}[H]
\centering
\includegraphics[width=0.9\textwidth]{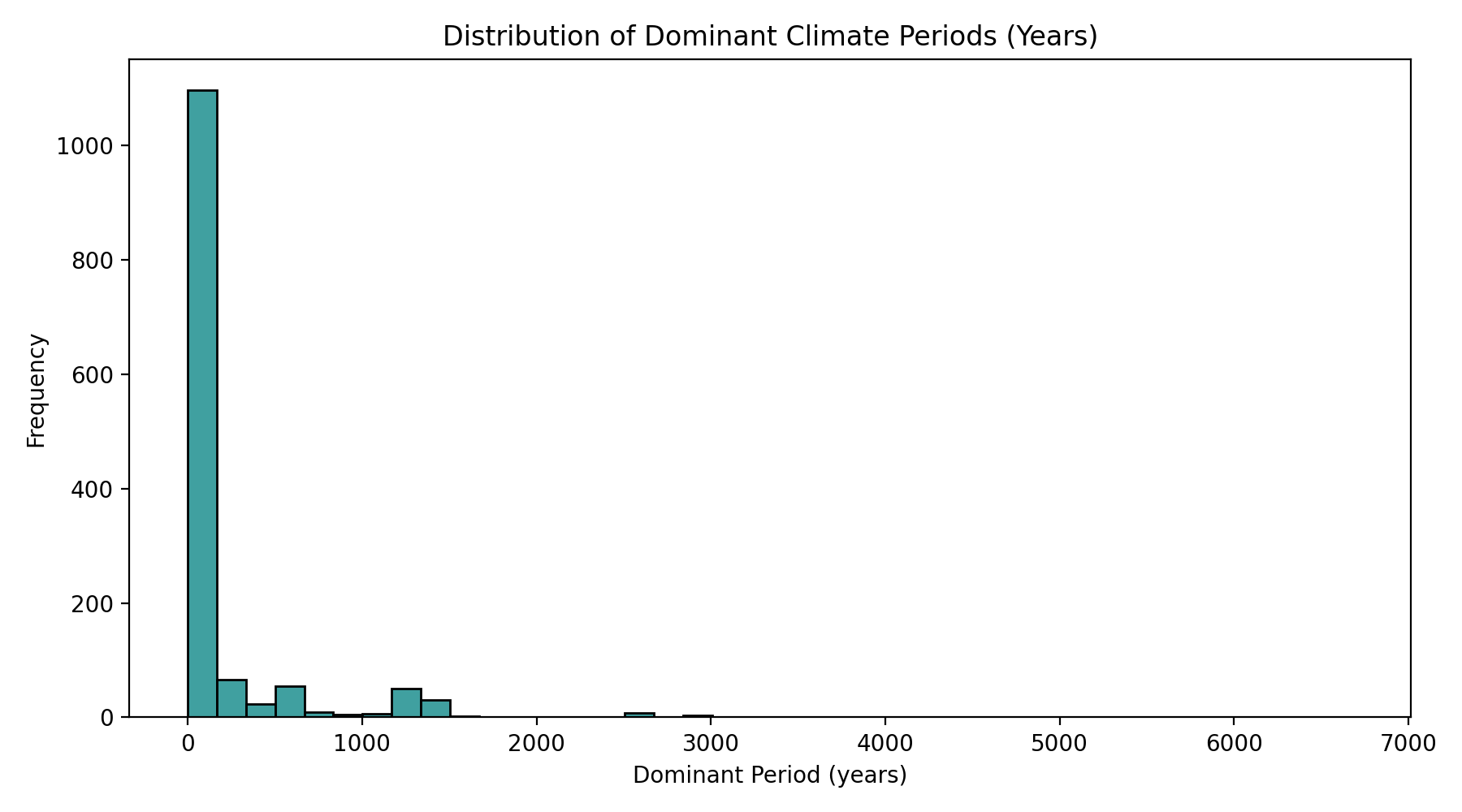}
\caption{Distribution of dominant periodicities across states and variables.}
\end{figure}

Dominant frequencies (Figure 7) cluster below 10 years, primarily representing ENSO and decadal oscillations. 
Longer apparent periods likely reflect statistical artifacts near boundary limits of the dataset.

\begin{figure}[H]
\centering
\includegraphics[width=0.9\textwidth]{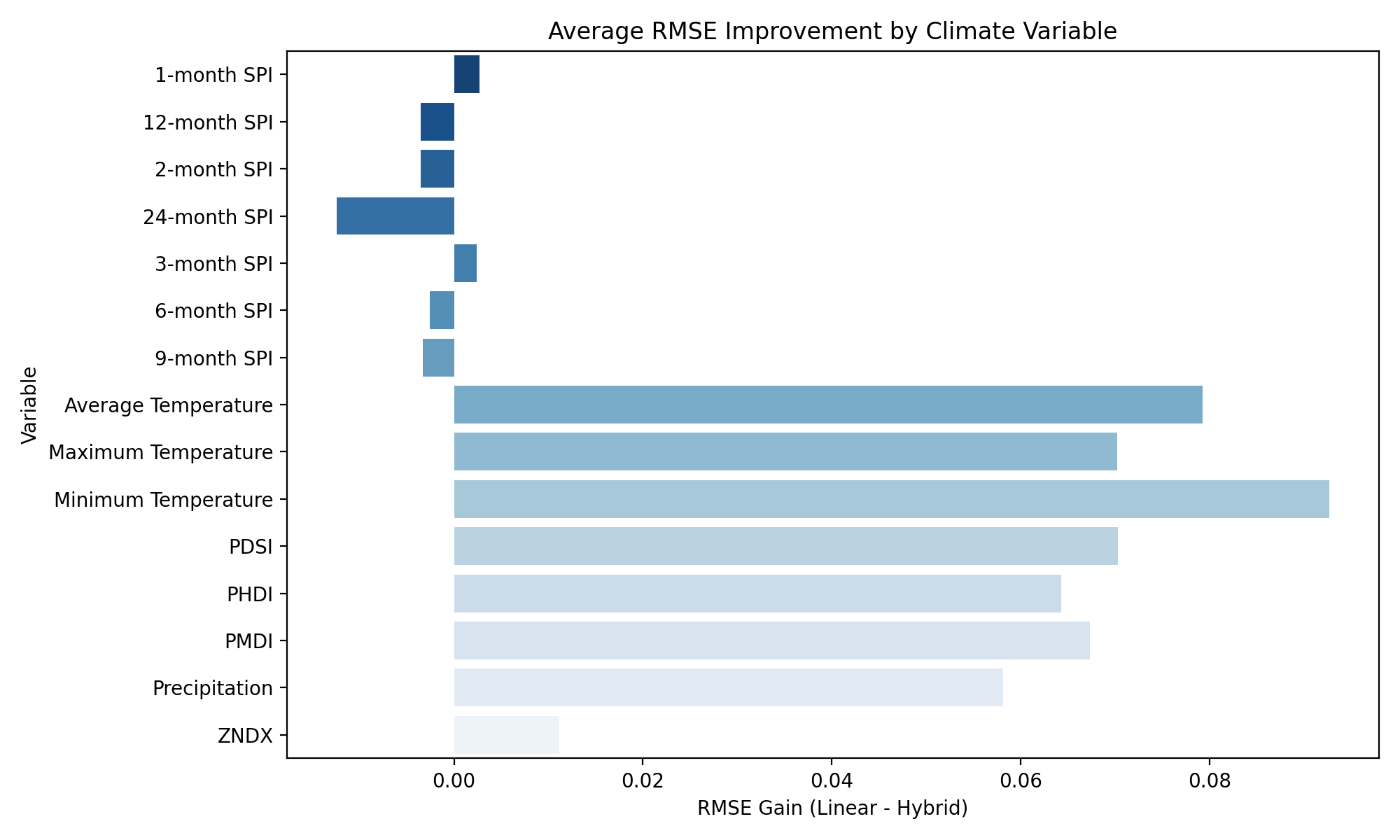}
\caption{Average RMSE improvement by variable.}
\end{figure}

Temperature-related variables show the greatest performance improvement (Figure 8) due to their stable seasonal periodicity. 
Precipitation and drought indices also benefit from harmonic terms but to a lesser degree, reflecting greater spatial and temporal heterogeneity in precipitation patterns.

\begin{figure}[H]
\centering
\includegraphics[width=0.95\textwidth]{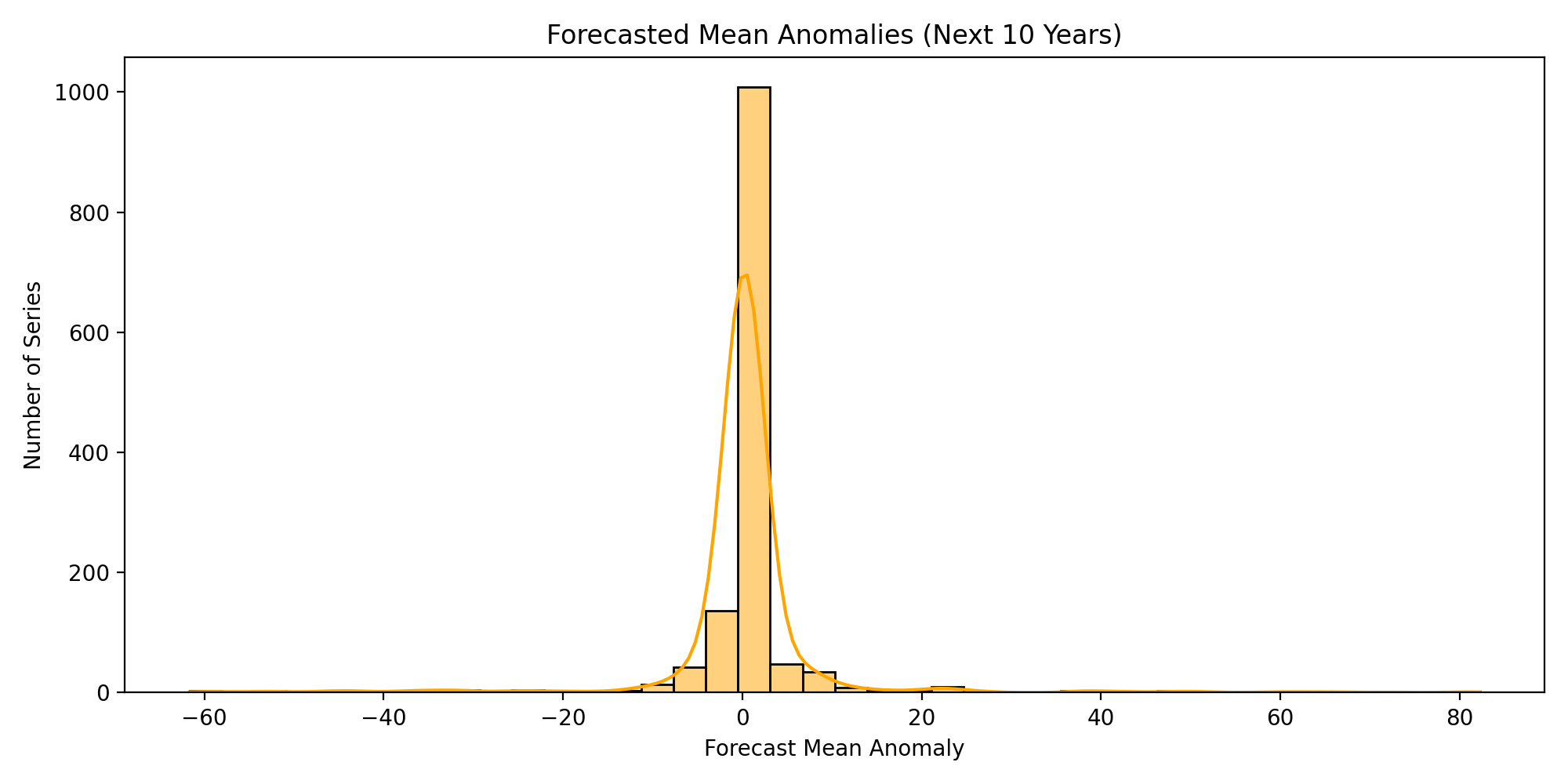}
\caption{Forecasted mean anomalies (2026–2035) using harmonic model.}
\end{figure}

Forecast distributions (Figure 9) center near zero, suggesting overall stationarity of anomalies in the near term. 
The slight positive skew indicates a potential increase in precipitation variability, consistent with intensifying hydrological extremes under warming conditions.

\begin{figure}[H]
\centering
\includegraphics[width=\textwidth]{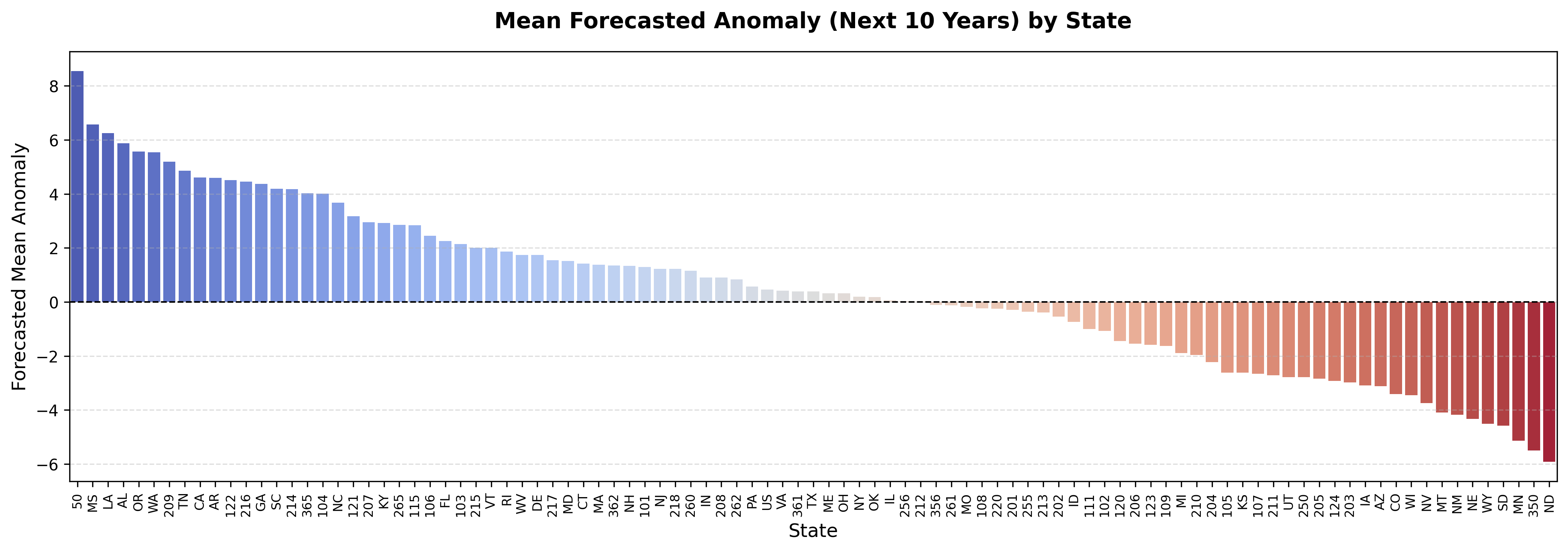}
\caption{Forecasted 10-year mean anomalies by state.}
\end{figure}

Figure 10 shows regionally differentiated forecasts: increased precipitation in the Gulf Coast and Pacific Northwest, and mild drying in the northern Great Plains. 
At the upper end of the distribution, states such as Louisiana and Mississippi (and, in the Northwest, Washington/Oregon) exhibit relatively positive projected anomalies.
At the lower end, North Dakota and Minnesota show relatively negative projected anomalies, consistent with northern Plains drying. 
Mid-range examples include large-population or transition states such as California and Texas, which cluster near small positive or near-zero values. 
These specific cases illustrate the broader spatial pattern and align with known ENSO teleconnections (enhanced wetness in the southern tier and reduced precipitation in parts of the northern tier).

\section{Conclusions}

This study applied harmonic and wavelet decomposition to more than a century of U.S. climate records to investigate temporal variability in temperature, precipitation, and drought indices. By decomposing monthly anomalies into frequency components and modeling both cyclic and monotonic behavior, we evaluated how periodic structure contributes to explaining and predicting observed climate variability. 
The results indicate that U.S. climate variability is best described by a combination of quasi-stationary oscillations and intermittent regime shifts rather than by a simple linear trend. Across most states and variables, incorporating harmonic terms derived from dominant Fourier frequencies substantially reduced model error relative to trend-only baselines, confirming that cyclical variability provides independent predictive value. 
Wavelet analysis further revealed that the strength of interannual oscillations—particularly in the 2–7~year range associated with El~Niño–Southern~Oscillation (ENSO)—has varied through time, peaking during the mid- to late 20th century. 
The detection of structural breaks corresponding to events such as the 1930s Dust Bowl and the 1976–1977 Pacific climate regime shift supports the view that abrupt transitions and periodic oscillations jointly shape U.S. hydroclimate behavior. Taken together, these findings highlight the limitations of interpreting climate evolution exclusively through linear trends. 
While long-term warming or wetting trends remain important indicators of change, the timing and amplitude of oscillatory modes determine many of the most societally relevant outcomes, including drought onset, flood risk, and agricultural productivity. 
Harmonic and wavelet methods therefore provide a complementary statistical framework that preserves both the long-term trajectory and the temporal structure of variability.

\subsection*{Limitations and Future Work}
Several methodological constraints should be noted. 
First, harmonic decomposition assumes that periodicities are approximately stationary and linearly additive, which may not fully capture nonlinear feedbacks or evolving phase relationships among coupled climate modes. Second, while wavelet transforms localize frequency variation in time, they have limited resolution for very low-frequency (multi-decadal) trends. 
Third, the regression models evaluated here are purely statistical and do not explicitly incorporate physical processes such as soil moisture feedbacks or atmospheric circulation patterns. Future work should address these limitations by integrating nonlinear spectral methods—such as empirical mode decomposition, singular spectrum analysis, or neural-network-based harmonic modeling—to better capture evolving periodicities and nonlinear phase shifts. 
Extending the framework to paleoclimate reconstructions or global reanalysis data could further reveal whether the observed quasi-periodic structure persists on centennial scales or under changing boundary conditions. In summary, harmonic and wavelet decomposition together provide an efficient, interpretable means of diagnosing the temporal organization of U.S. climate variability. By identifying stable oscillatory modes and their temporal evolution, these methods improve predictive performance and offer a more physically grounded interpretation of how variability, rather than solely trend, defines the changing character of the U.S. climate system.

\section*{Acknowledgments}
Data were obtained from NOAA’s National Centers for Environmental Information (NCEI) nClimDiv database. 
Computational analyses employed \texttt{pandas}, \texttt{numpy}, \texttt{matplotlib}, \texttt{pywt}, and \texttt{ruptures} libraries.

\end{document}